\documentclass[reprint]{revtex4-2}

\usepackage{graphicx}
\usepackage{amssymb}   
\usepackage{amsmath}    
\usepackage{graphicx}   
\usepackage{verbatim}   
\usepackage{color}      
\usepackage{hyperref}   
\usepackage{bm} 
\usepackage{ulem}

\usepackage{titlesec} 
\titleformat{\section}{\bfseries\MakeUppercase}{}{0pt}{}
\titlespacing*{\section}{0pt}{3ex}{0.3ex}
\titleformat{\subsection}{\bfseries}{}{0pt}{}
\titlespacing*{\subsection}{0pt}{3ex}{0.3ex}

\makeatletter
\renewcommand{\fnum@figure}{\textbf{Fig. \thefigure}}
\makeatother

\begin{document}

\preprint{}

\title{Electric-field-driven magnetic switching and tightly bound interlayer excitons in bilayer CrSBr}

\author{Xiubin Li$^{1}$, Yu Sun$^{1}$, Xuanji Wang$^{2}$, Chunyan Wang$^{1}$  Kenji Watanabe$^{3}$, Takashi Taniguchi$^{3}$, Sheng Liu$^{2}$,\\ Ting Yu$^{2}$, Liang Li$^{1*}$, Tao Zhang$^{1*}$, Jing Li$^{1*}$}

\affiliation{$^1$Wuhan National High Magnetic Field Center and School of Physics, Huazhong University of Science and Technology, Wuhan 430074, China
\\$^2$School of Physics and Technology, Wuhan University, Wuhan 430072, China
\\$^3$National Institute for Material Science, 1-1 Namiki, Tsukuba 305-0044, Japan}

\begin{abstract}
Electric field control of magnetic order in two-dimensional (2D) van der Waals magnets is a central goal for low-power spin-based technologies. In the ambient-stable antiferromagnet CrSBr, strong magnetic anisotropy and robust exciton-spin coupling provide a favorable platform, yet deterministic electric field control of its magnetic phases has not been achieved. Here we demonstrate electric-field-driven reversible switching between antiferromagnetic and ferromagnetic states in dual-gated bilayer CrSBr without intentional carrier doping. In parallel, photoluminescence measurements resolve a tightly bound interlayer exciton with an intrinsic dipole moment of only $\sim$1 e\AA. The electric field dependence of the magnetic phase transition reveals two coexisting mechanisms: a linear magnetoelectric effect in the antiferromagnetic state and an electric-field-modulated interlayer exchange coupling. Their interplay accounts for the asymmetric evolution of the critical magnetic field. Our results establish bilayer CrSBr as a promising 2D material for electrically controlled spin-optoelectronic functionalities.
\end{abstract}

\maketitle

\noindent
The capability to manipulate magnetic order using electric fields is of central importance for both fundamental condensed matter physics and next-generation spintronic technologies \cite{Matsukura2015}. Electric field control provides a low-dissipation route to tune magnetic phase transitions \cite{Eerenstein2006,Chiba2011}, while simultaneously offering opportunities for electrically controlled non-volatile memory and multifunctional spin-optoelectronic devices \cite{Prinz1998,Chu2008,Matsukura2015}. Over the past decades, electrically controlled magnetism has been extensively investigated in a wide range of bulk material systems, including ferromagnetic (FM) metals \cite{Weisheit2007,Maruyama2009,Wang2012}, dilute magnetic semiconductors \cite{Ohno2000,Dietl2000,Macdonald2005,Chiba2008,Matsukura2015}, and multiferroic compounds \cite{Chu2008,Wu2010,Heron2011}. The emergence of two-dimensional (2D) van der Waals (vdW) magnetic materials, initiated by the discoveries of long-range magnetic order in CrI$_3$ \cite{Huang2017} and Cr$_2$Ge$_2$Te$_6$ \cite{Gong2017}, has established a new platform for magnetoelectric (ME) functionalities at the atomic limit. In particular, electrically controlled switching between antiferromagnetic (AFM) and FM states was realized in bilayer CrI$_3$ \cite{Jiang2018,Huang2018,Jiang2018_2}, highlighting the remarkable potential of 2D magnets for electric field manipulation of magnetic phases. Extending such capabilities to a broader family of ambient-stable magnetic semiconductors with strong optical responses remains highly desirable.

Among recently discovered vdW magnetic semiconductors, CrSBr has attracted substantial attention owing to its pronounced magnetic anisotropy \cite{Telford2020,Wilson2021,Lee2021,Sun2025}, robust exciton-magnon coupling \cite{Bae2022,Diederich2022,Datta_2025,Diederich2025,Dirnberger2026}, strong spin-dependent optical responses \cite{Ziegler2025,Qin2026,Wang2026}, and excellent ambient stability \cite{Ziebel2024}. Few-layer CrSBr furthermore exhibits weak interlayer magnetic coupling and strong charge-spin correlations, enabling doping-induced AFM-FM transitions \cite{Xie2023,Tabataba-Vakili2024,Zhao2025}. These properties render CrSBr an appealing platform for electrically tunable magnetism and spin-optoelectronic functionalities. Nevertheless, deterministic electric field control of magnetic order in CrSBr remains unexplored. Beyond the possibility of a linear ME effect allowed by time- and spatial-inversion symmetry breaking \cite{Sivadas2016,Jiang2018}, recent theoretical studies also suggest that electric fields could directly modify the weak interlayer magnetic coupling in bilayer CrSBr, providing an additional route for tuning magnetic phase stability \cite{Wang2023}.

Crucially, the microscopic origin of excitonic states in CrSBr remains under active debate. Several competing interpretations have been proposed, including assignments of $X_{\mathrm{a}}$ and $X_{\mathrm{b}}$ to transitions involving a common valence band but distinct conduction bands \cite{Wilson2021,Tabataba-Vakili2024,Krelle2025}, surface and bulk excitons arising from identical interband transitions but modified dielectric screening \cite{Shao2025}, and more recently unipolar charged excitons emerging exclusively in the electron-doped regime \cite{Tabataba-Vakili2024}. Furthermore, whether bilayer CrSBr hosts interlayer excitons remains experimentally unresolved, despite their central role in many vdW semiconductor structures \cite{Wilson2021_1,zheng2022,Zhu2025}.

Here, we investigate the evolution of excitonic states and magnetic order in dual-gated bilayer CrSBr devices under electric field control. Notably, the CrSBr channel is electrically floating without direct grounding contact, enabling electric field modulation in the absence of intentional carrier injection. Low-temperature photoluminescence (PL) spectroscopy reveals electrically tunable intralayer and interlayer excitonic states. We observe pronounced electric-field-induced redistribution of oscillator strength between intralayer excitonic species, supporting the assignment of $X_{\mathrm{a}}$ and $X_{\mathrm{b}}$ to transitions involving distinct conduction bands. In addition, we identify an interlayer exciton $X_{\mathrm{i}}$ exhibiting a remarkably small intrinsic dipole moment of approximately 1 e\AA, indicative of strong interlayer hybridization and tightly bound excitonic character. Most importantly, we demonstrate electric-field-driven reversible AFM-FM switching in bilayer CrSBr. The electric field dependence of the critical magnetic field reveals the coexistence of linear ME effects and electric-field-modulated interlayer magnetic coupling. These results establish bilayer CrSBr as a versatile platform in which excitonic and magnetic degrees of freedom can be simultaneously manipulated by electric fields.

\vspace{-\parskip}
\section{Results and Discussion}
\vspace{-2\parskip}

\begin{figure*}[!th]
    \centering
    \includegraphics[scale=0.53]{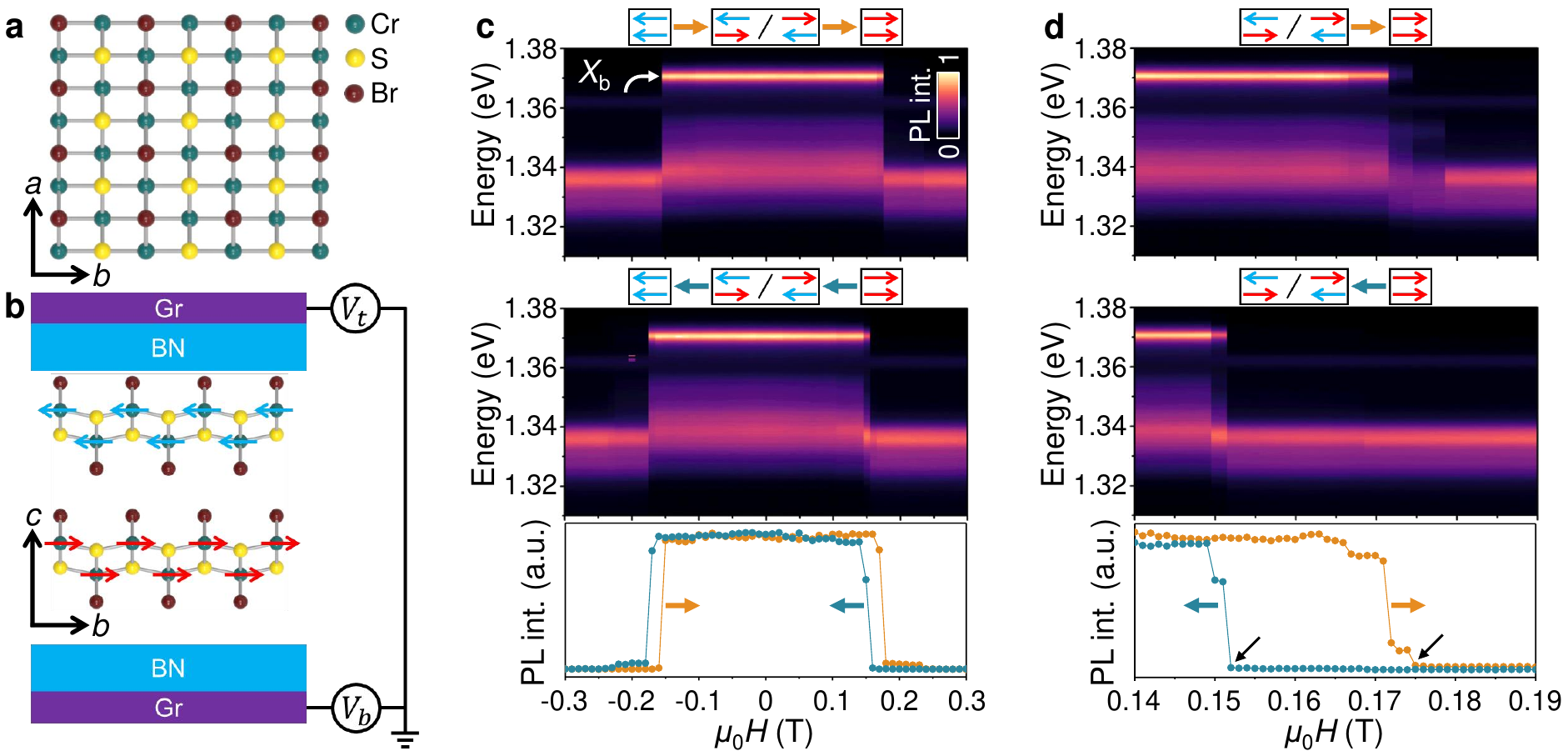}
    \caption{\textbf{Crystal structure and magneto-PL spectra of bilayer CrSBr.} \textbf{a} Top view of the crystal structure of bilayer CrSBr, with arrows indicating the crystallographic axes. \textbf{b} Schematic illustration of the dual-gated device structure, and the side view of CrSBr crystal structure. Bilayer CrSBr is encapsulated by two hBN flakes, while few-layer graphite serves as the top and bottom gates. $V_{\mathrm{t}}$ and $V_{\mathrm{b}}$ denote the top- and bottom-gate voltages, respectively. Blue and red arrows represent opposite magnetic-moment orientations. \textbf{c} Upper two panels: magneto-PL spectra measured under opposite magnetic-field sweeping directions along the $b$ axis. The schematics above the panels illustrate the evolution of magnetic-moment orientations and the corresponding magnetic-field sweep directions (orange and cyan arrows indicate opposite sweep directions). Bottom panel: extracted PL intensity near 1.371 eV. \textbf{d} Same as \textbf{c}, but measured over a smaller magnetic-field range with a finer field step of 1 mT. Black arrows indicate the critical magnetic fields associated with the spin-flip transition.}
    \label{Figure_1}
\end{figure*}

Dual-gated hBN-encapsulated bilayer CrSBr devices were fabricated to investigate the influence of out-of-plane electric fields on excitonic states and magnetic order (see Methods). The crystal structure of bilayer CrSBr is shown in Figs. 1a and 1b. CrSBr crystallizes in an orthorhombic structure with space group $P_{\mathrm{mmn}}$ ($D_{\mathrm{2h}}$). The magnetic moments align preferentially along the crystallographic $b$ axis, corresponding to the magnetic easy axis, while the $a$ and $c$ axes serve as the intermediate and hard magnetic axes, respectively \cite{Wilson2021}. Optical images and device schematics are presented in Figs. 1b and S1. Bilayer CrSBr is encapsulated by two hBN flakes, while few-layer graphite flakes serve as electrostatic gates. The CrSBr channel is intentionally left electrically floating (no direct electrical contact), allowing a large displacement field to be applied along the $c$ axis without external carrier injection into the bilayer system. This device geometry enables direct investigation of ME coupling and electric-field-induced excitonic reconstruction.

We first discuss the correlation between excitonic states and magnetic order in bilayer CrSBr. Figure 1c presents magneto-PL measurements performed at 3.7 K, well below the Néel temperature of approximately 140 K \cite{Lee2021}, with the magnetic field applied along the crystallographic $b$ axis. The upper panel displays PL spectra measured while sweeping the magnetic field from $-0.3$ T to $+0.3$ T. Abrupt spectral changes emerge near $-0.15$ T and $+0.18$ T, corresponding to spin-flip transitions between AFM and FM states \cite{Tabataba-Vakili2024,Wu2025}. The schematics above the panels illustrate the evolution of magnetic configurations during the magnetic-field sweep. A pronounced modulation of the highest-energy excitonic feature accompanies these magnetic phase transitions. This excitonic state, denoted as $X_{\mathrm{b}}$, appears at approximately 1.371 eV in the AFM state and 1.362 eV in the FM state, and has previously been assigned to an intralayer transition between the valence-band maximum (VBM) and the second conduction-band minimum (CBM2) \cite{Wilson2021,Tabataba-Vakili2024,Krelle2025,Wu2025}. The strong suppression of $X_{\mathrm{b}}$ intensity in the FM state originates from restoration of spatial inversion symmetry, which transforms the VBM-CBM2 transition from dipole allowed in the AFM state to dipole forbidden in the FM state \cite{Wu2025}. The middle panel of Fig. 1c displays PL spectra measured during the reverse magnetic-field sweep from $+0.3$ T to $-0.3$ T, again revealing abrupt magnetic phase transitions. To more clearly visualize the hysteretic behavior, the resonance intensity near 1.371 eV is extracted and plotted in the lower panel. The resulting hysteresis loop demonstrates the first-order nature of the magnetic phase transition.

\begin{figure*}[!t]
    \centering
    \includegraphics[scale=0.53]{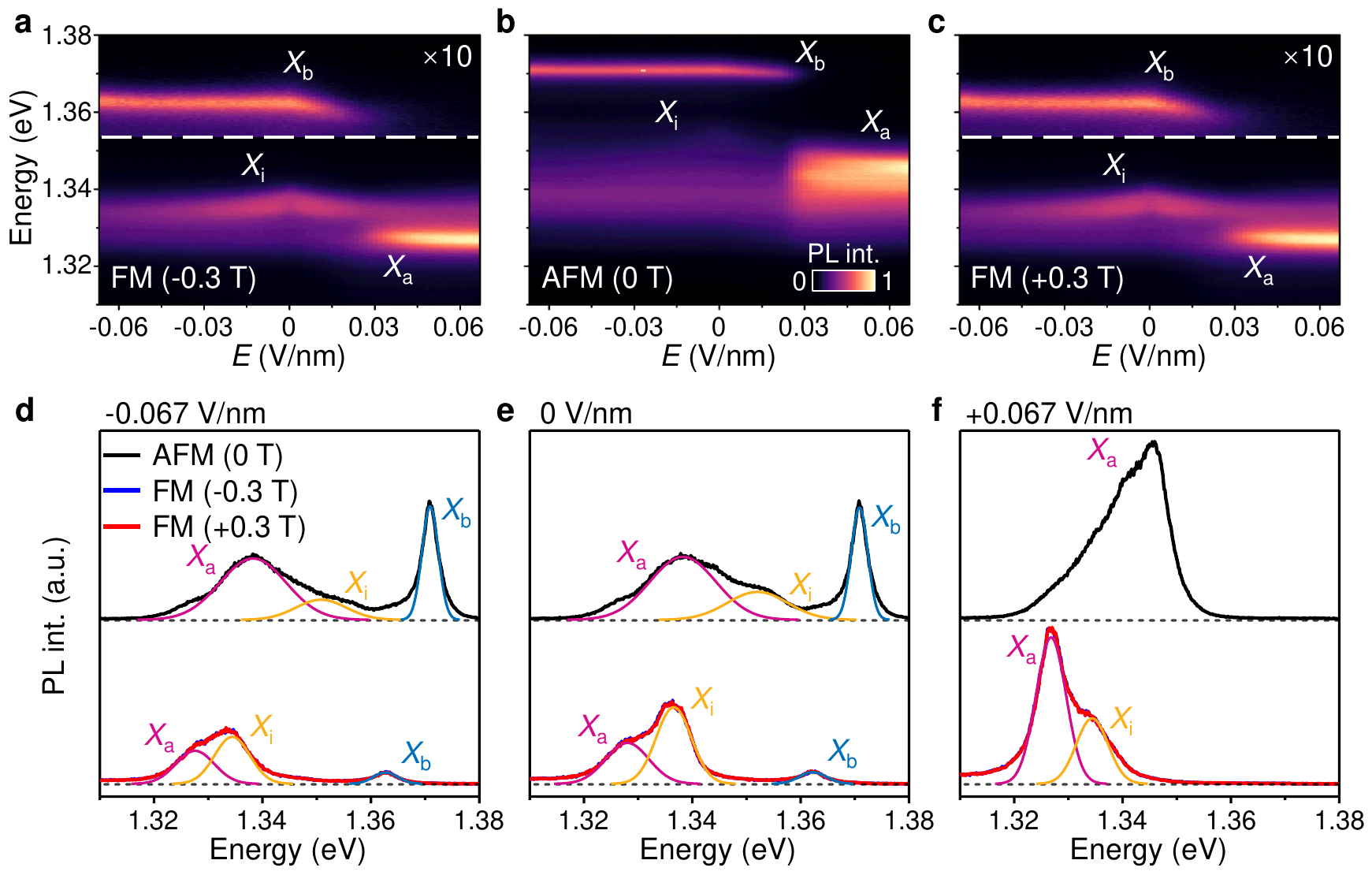}
    \caption{\textbf{Electric-field-tunable and magnetic-order-dependent excitonic states in bilayer CrSBr.} \textbf{a-c} Electric-field-dependent PL spectra measured at $-0.3$, 0, and $+0.3$ T, respectively. To clearly visualize $X_{\mathrm{b}}$ in the FM state ($\pm0.3$ T), the PL intensity in the high-energy region of \textbf{a} and \textbf{c} is magnified by a factor of 10. \textbf{d-f} PL spectra measured under different magnetic states at $-0.067$, 0, and $+0.067$ V/nm, respectively. Magenta, yellow, and royal-blue curves represent Gaussian fitting components corresponding to $X_{\mathrm{a}}$, $X_{\mathrm{i}}$, and $X_{\mathrm{b}}$, respectively. In panel \textbf{f}, the complicated spectral profile of $X_{\mathrm{a}}$ prevents reliable Gaussian fitting.}
    \label{Figure_2}
\end{figure*}

To determine the critical magnetic fields $\mu_{0}H_{\mathrm{C}}$ ($\mu_{0}$ is the vacuum permeability) with higher precision, magneto-PL measurements were performed using magnetic-field increments of 1 mT, as shown in Fig.~1d. Here, $\mu_{0}H_{\mathrm{C}}$ is defined as the magnetic field at which the resonance intensity begins to rise sharply in the FM state, as indicated by the black arrows. The critical fields for the AFM--FM transition are approximately $+0.175$ T and $-0.176$ T (Fig.~S2), comparable to previous reports on hBN-encapsulated CrSBr \cite{Boix-Constant2022,Tabataba-Vakili2024,Wu2025,Dong2026,Chen2026} but consistently larger than those observed in as-exfoliated flakes \cite{Wilson2021,Shi2025}. In addition, the hysteresis width reaches approximately 23 mT, in agreement with earlier studies of encapsulated devices \cite{Tabataba-Vakili2024} and substantially exceeding that of unencapsulated samples \cite{Wilson2021}. These enhancements are likely associated with dielectric-environment effects that modify the magnetic anisotropy and its balance with interlayer AFM exchange interactions \cite{Jacobs1947,Rudenko2023,Boix-Constant2025}. Similar encapsulation-dependent behavior is also observed in Device 2 (Fig.~S3) and in bilayer CrI$_3$ \cite{Huang2017,Jiang2018}. Collectively, the elevated critical fields and broadened hysteresis indicate robust magnetic anisotropy in hBN-encapsulated bilayer CrSBr, establishing a favorable platform for investigating electric field control of magnetism \cite{Chang2023}.

We next investigate the evolution of excitonic states under an out-of-plane electric field $E$ applied along the crystallographic $c$ axis, where the positive field direction is defined from the top layer toward the bottom layer. Electric-field-dependent PL spectra were measured under magnetic fields sufficiently far from $\mu_{0}H_{\mathrm{C}}$ to ensure stable AFM or FM configurations. As shown in Figs. 2a–c, three distinct excitonic features are resolved. The lowest-energy feature is denoted as $X_{\mathrm{a}}$, appearing near 1.339 eV in the AFM state and 1.328 eV in the FM state at $E=0$. Consistent with previous studies, $X_{\mathrm{a}}$ is assigned to an intralayer transition between the VBM and the first conduction-band minimum (CBM1) \cite{Wilson2021,Tabataba-Vakili2024,Wu2025,Shi2025,Krelle2025}.

A clear electric-field-induced redistribution of oscillator strength occurs between $X_{\mathrm{a}}$ and $X_{\mathrm{b}}$. As shown in Figs. 2a–c, when $E$ increases from 0 to approximately $+0.03$ V/nm, the resonance intensity of $X_{\mathrm{a}}$ gradually increases, while that of $X_{\mathrm{b}}$ decreases rapidly. Beyond approximately $+0.03$ V/nm, $X_{\mathrm{a}}$ becomes strongly enhanced and $X_{\mathrm{b}}$ is almost completely quenched, indicating substantial reconstruction of the excitonic landscape. By comparison, under negative electric fields, both the energy and intensity of $X_{\mathrm{a}}$ and $X_{\mathrm{b}}$ remain apparently unchanged. This pronounced asymmetry with respect to electric field direction suggests the involvement of localized states near the bottom CrSBr layer, possibly introduced during fabrication through temporary contact with the $\mathrm{SiO_2}$ substrate and the delamination process before hBN encapsulation. As illustrated schematically in Fig. S4, positive electric fields raise these localized states relative to the Fermi level, thereby suppressing transitions involving CBM2 and strongly reducing the oscillator strength of $X_{\mathrm{b}}$. Negative electric fields, by contrast, do not induce comparable unoccupied localized states and therefore impose much weaker influence on the intralayer excitonic transitions. This behavior further supports the interpretation of $X_{\mathrm{a}}$ and $X_{\mathrm{b}}$ as transitions involving distinct conduction bands.

In addition to the intralayer excitons, the electronic structure of bilayer CrSBr also permits an interlayer transition between the VBM and CBM1, as illustrated in Fig. S4a. Such a transition can give rise to an interlayer exciton, denoted as $X_{\mathrm{i}}$, in which the electron and hole reside predominantly in different layers. Compared with intralayer excitons, interlayer excitons generally possess reduced binding energies owing to the extra vertical spatial separation between the electron-hole pair. At zero electric field, $X_{\mathrm{i}}$ and $X_{\mathrm{a}}$ involve analogous VBM-CBM1 transitions with similar interband hopping energies. The reduced binding energy of the interlayer exciton therefore leads to a slightly higher PL emission energy. Upon application of an out-of-plane electric field, the bilayer band structure evolves toward a type-II band alignment, reducing the effective interlayer hopping energy, as schematically illustrated in Figs. S4b and S4c. Notably, this field-induced type-II alignment depends primarily on the magnitude rather than the direction of the electric field and is insensitive to localized states. Consequently, the electric field evolution of the interlayer exciton is expected to depend predominantly on $|E|$.

Besides $X_{\mathrm{a}}$ and $X_{\mathrm{b}}$, a third excitonic feature is observed in Figs. 2a–c near 1.354 eV in the AFM state and 1.337 eV in the FM state at $E=0$. This excitonic state exhibits several characteristics expected for an interlayer exciton, including a symmetric electric field dependence with respect to field direction and a pronounced Stark-like energy shift governed primarily by $|E|$. We therefore attribute this feature to an interlayer exciton $X_{\mathrm{i}}$. As $|E|$ increases, the resonance energy of $X_{\mathrm{i}}$ continuously redshifts before gradually approaching saturation at larger electric fields (Fig. S5). At relatively small electric fields ($|E|<0.011$ V/nm), the energy shift of $X_{\mathrm{i}}$ varies approximately linearly with $|E|$, consistent with a first-order Stark-like response arising from electric-field-induced reduction of interlayer hopping energy. From the initial slope, the intrinsic dipole moment of $X_{\mathrm{i}}$ is estimated to be approximately 1 e\AA\ (Fig. S5), substantially smaller than typical values reported for interlayer excitons in transition-metal dichalcogenide bilayers ($\sim$ 5 e\AA) \cite{Ciarrocchi2019,Leisgang2020,Barre2022}. This unusually small dipole moment indicates that the electron and hole remain strongly hybridized across the two layers despite their interlayer character. This observation is consistent with the reported tightly bound intralayer excitonic character in CrSBr \cite{Smiertka2026}. At larger electric fields, increased electron-hole separation reduces both the exciton binding energy and wavefunction overlap, leading to gradual saturation of the energy shift together with suppression of PL intensity.

\begin{figure*}[ht]
    \centering
    \includegraphics[scale=0.54]{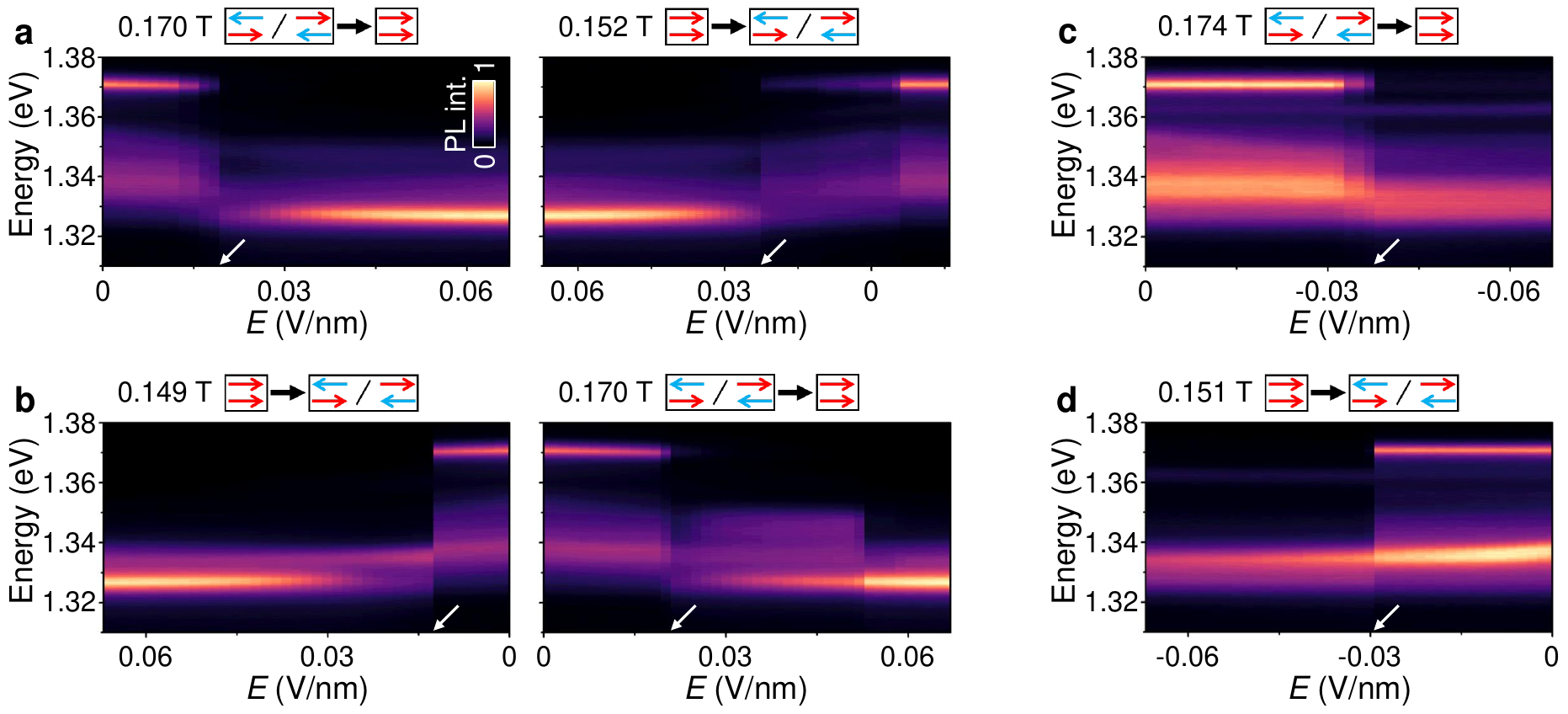}
    \caption{\textbf{Electric-field-driven magnetic switching in bilayer CrSBr.} \textbf{a-d} PL spectra as a function of electric field $E$ measured under constant magnetic fields. The $E$-sweeping direction is defined by the left-to-right orientation along the horizontal axis. The schematics above the PL maps illustrate the corresponding magnetic-field configurations and the electric-field-driven magnetic phase-transition processes. The PL spectra shown in the left and right panels of \textbf{a} and \textbf{b} were measured sequentially, separated by a magnetic-field adjustment process performed at fixed $E$. White arrows indicate the critical magnetic fields associated with magnetic phase transitions.}
    \label{Figure_3}
\end{figure*}

We next discuss the influence of magnetic order on the three excitonic states. Figures 2d–f present PL spectra measured under AFM and FM configurations at representative electric fields. The interlayer transition associated with $X_{\mathrm{i}}$ is strongly influenced by magnetic order. In the FM state, the interlayer transition between the VBM and CBM1 remains spin allowed, whereas in the AFM state it becomes partially spin blocked due to opposite spin configurations in adjacent layers. Nevertheless, strong exciton-magnon coupling in CrSBr enables weak spin-forbidden interlayer transitions mediated by magnons \cite{Bae2022,Diederich2022,Datta_2025,Diederich2025,Dirnberger2026}, preventing complete quenching of $X_{\mathrm{i}}$ in the AFM state. Consequently, the PL intensity of $X_{\mathrm{i}}$ remains finite but substantially weaker in the AFM configuration than in the FM configuration. Moreover, because $X_{\mathrm{i}}$ and $X_{\mathrm{a}}$ involve analogous VBM-CBM1 transitions at $E=0$, their resonance intensities exhibit a complementary relationship. Enhancement of $X_{\mathrm{i}}$ in the FM state is accompanied by suppression of $X_{\mathrm{a}}$, reflecting redistribution of oscillator strength between interlayer and intralayer channels. Similar behavior is also observed in Device 2 (Fig. S6), demonstrating the reproducibility of the phenomenon. The strong sensitivity of the excitonic structure to both electric field and magnetic order suggests that excitonic reconstruction and ME coupling are strongly correlated in bilayer CrSBr.

Having established the strong coupling between excitonic and magnetic degrees of freedom, we next investigate whether magnetic order itself can be directly controlled by electric fields. Figure 3 presents electric-field-dependent PL spectra measured under magnetic fields close to $\mu_{0}H_{\mathrm{C}}$. As shown in Fig. 3a, successive AFM-FM-AFM transitions can be reversibly induced by positive electric fields. Starting from the AFM state at 0.170 T, the AFM-FM transition occurs near $+0.019$ V/nm as the electric field increases. The FM state remains stable over a broad electric-field range and can be switched back to the AFM state upon lowering magnetic field to 0.152 T and reversing the electric-field sweep, completing a fully reversible AFM-FM-AFM cycle. Figure 3b displays an analogous sequence of reversible FM-AFM-FM transitions induced by positive electric fields. Similar behavior is also observed for opposite magnetic-field directions with nearly identical switching characteristics (Fig. S7), confirming the deterministic and reversible nature of the electric-field-driven magnetic switching.

In contrast, negative electric fields induce markedly different switching behavior. As shown in Figs. 3c and 3d, negative electric fields can separately drive AFM-FM and FM-AFM transitions. However, these two switching processes cannot be connected into a continuous electric-field-controlled cycle. Re-establishing the initial magnetic state (i.e. FM phase in Fig. 3d) requires additional magnetic-field-driven transitions, indicating that the switching process is not fully reversible. Positive electric fields therefore impose a substantially stronger influence on the magnetic free-energy landscape than negative electric fields. Similar asymmetric switching behavior is also observed in Device 2 (Figs. S8 and S9), demonstrating that the phenomenon is intrinsic rather than device specific.

The reversible magnetic switching shown in Fig. 3 suggests that the applied electric field modulates the free-energy balance between the AFM and FM states. The shift of $\mu_{0}H_{\mathrm{C}}$ directly reflects changes in the free-energy difference between the AFM and FM states. Figure 4a summarizes the electric field dependence of $\mu_{0}H_{\mathrm{C}}$ for the AFM-FM transition, extracted from high-resolution magneto-optical measurements performed with magnetic-field steps as small as 0.2 mT following the application of a fixed electric field at zero magnetic field (representative measurements are shown in Figs. S10a-e). Three observations are particularly noteworthy. $\mu_{0}H_{\mathrm{C}}$ decreases systematically with increasing $|E|$. The magnitude of the reduction depends strongly on electric field direction, with positive electric fields producing a substantially larger decrease in $\mu_{0}H_{\mathrm{C}}$ than negative electric fields. In addition, for positive electric fields, the variation of $\mu_{0}H_{\mathrm{C}}$ becomes strongly suppressed once $E$ exceeds approximately $+0.029$ V/nm, resulting in a pronounced turning point. Similar trends are observed for the FM-AFM transition and are reproduced in Device 2 (Figs. S11 and S12).

The asymmetric modulation of $\mu_{0}H_{\mathrm{C}}$ can be quantitatively understood by considering two distinct electric-field-dependent contributions to the free-energy difference between the AFM and FM phases (Fig. 4b). In bilayer CrSBr, the AFM state simultaneously breaks time-reversal and spatial-inversion symmetries, thereby permitting a linear ME effect \cite{Sivadas2016,Jiang2018}. By contrast, spatial inversion symmetry is restored in the FM state, suppressing the linear ME contribution. Consequently, the free energies of the AFM and FM states can be expressed as $F_{\mathrm{AFM}}=2F_{0}-J+F_{\mathrm{ME}}$ and $F_{\mathrm{FM}}=2F_{0}+J-\mu_{0}M_{0}H$, respectively. Here, $F_{0}$ denotes the free energy of an individual CrSBr layer, $J$ is the interlayer exchange-coupling energy, and $M_{0}$ is the saturation magnetization of bilayer CrSBr in the FM state. The linear ME effect contributes an additional term $F_{\mathrm{ME}}=-\alpha_{cb}EH$, corresponding to modulation of the magnetization along the $b$ axis by an electric field applied along the $c$ axis \cite{Fiebig2005}. Here, $\alpha_{cb}$ represents the corresponding off-diagonal component of the ME susceptibility tensor.

\begin{figure*}[!t]
    \centering
    \includegraphics[scale=0.52]{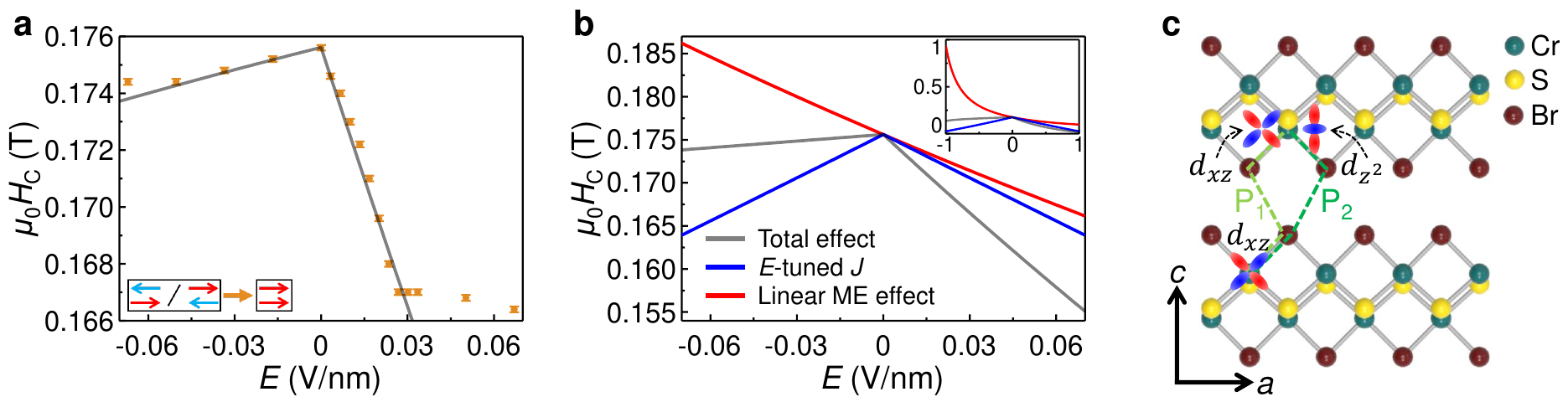}
    \caption{\textbf{The mechanism of electric-field-driven magnetic switching in bilayer CrSBr.} \textbf{a} Electric-field-dependent $\mu_{0}H_{\mathrm{C}}$ for the AFM--FM transitions in Device 1. The gray curve plots the fitting results, which combines a linear ME effect and $E$-tuned interlayer magnetic coupling mechanism. \textbf{b} Comparison of contributions to $\mu_{0}H_{\mathrm{C}}$-modulation from the linear ME effect and $E$-tuned interlayer magnetic coupling. The gray line is the fitting curve extracted from panel \textbf{a}. The contributions from linear ME effect and $E$-tuned interlayer magnetic coupling are represented by the red and blue curves, respectively. Inset shows the fitting results in an expanded electirc field range. \textbf{c} Side view of the crystal structure of bilayer CrSBr and schematic illustration of the super-super-exchange paths for interlayer magnetic coupling. The bright and dark green dashed lines mark the two super-super-exchange paths of $\mathrm{P}_{1}$ and $\mathrm{P}_{2}$, respectively.}
    \label{Figure_4}
\end{figure*}

In addition to the linear ME effect, the interlayer exchange coupling $J$ itself is dependent on electric field. Recent theoretical work predicts that electric-field-induced modifications of $J$ can even drive magnetic order switching at zero magnetic field when sufficiently large electric fields are applied \cite{Wang2023}. As illustrated in Fig. 4c, $J$ is governed by two competing super-super-exchange pathways. The $\mathrm{P}_{1}$ pathway involves interlayer hopping between two occupied $d_{\mathrm{xz}}$ orbitals on neighboring layers of Cr atoms. Because this process only involves electrons carrying opposite spins, it predominantly contributes to interlayer AFM coupling. In contrast, the $\mathrm{P}_{2}$ pathway involves hopping between occupied $d_{\mathrm{xz}}$ and unoccupied $d_{\mathrm{z^{2}}}$ orbitals. This process contributes to both AFM and FM exchanges but preferentially favors FM coupling following Hund's rule. Nevertheless, both exchange pathways are sensitive to the energy differences between the participating Cr-3$d$ orbitals residing on adjacent layers, which are directly tunable by an electric field applied along the $c$ axis (see Supplementary Note 1 for detailed calculations). The resulting modulation of $J$ depends only on the magnitude of the electric field rather than its direction, giving rise to a symmetric contribution to the electric field tuning of the magnetic free energy.

The symmetric modification of $J$ alone cannot account for the clear asymmetry observed in Fig. 4a. The remaining asymmetry is naturally captured by the linear ME effect, which changes sign of $F_\mathrm{ME}$ upon reversing the electric field. By combining the linear ME effect and the electric-field-dependent interlayer exchange coupling, the critical magnetic field can be expressed as $\mu_{0}H_{\mathrm{C}}=\frac{2\mu_{0}J(|E|)}{\mu_{0}M_{0}-\alpha_{cb}E}$. This expression quantitatively captures the asymmetric evolution of $\mu_{0}H_{\mathrm{C}}$, and the corresponding fitting results are shown as gray curves in Figs. 4a and 4b (see Supplementary Note 1 for fitting details). The fit yields a linear volumetric ME coefficient of approximately 82 ps/m for bilayer CrSBr, comparable to the diagonal component of the linear ME susceptibility reported for bilayer CrI$_3$ \cite{Jiang2018} and substantially larger than values reported for typical conventional bulk ME materials \cite{Vaknin2004,Rivera2009,Lin2022,Xu2022,Xu2023}. A more recent report of out-of-plane magnetic-field-driven in-plane electric polarization switching in bilayer CrSBr similarly support the existence of a finite $\alpha_{cb}$ \cite{Wang2025}. Multiple microscopic mechanisms may contribute to the observed linear ME effect, including electric-field-driven interlayer charge redistribution \cite{Jiang2018}, electric-field-modified $p$-$d$ hybridization \cite{Murakawa2010,Luo2017}, and the Katsura-Nagaosa-Balatsky mechanism \cite{Wang2025}.

The fitting further provides an estimate of the energy scale associated with the electric-field-dependent interlayer exchange coupling, corresponding to a reduction of $J$ by approximately 9.47 $\mu$J/m$^{2}$ per V/nm. The linear ME effect and the electric-field-induced modulation of interlayer exchange contribute comparably to the evolution of $\mu_{0}H_{\mathrm{C}}$, as illustrated in Fig. 4b. The red and blue curves represent the effective changes in $\mu_{0}H_{\mathrm{C}}$ arising solely from the linear ME effect and the electric-field-dependent exchange coupling, respectively. Positive electric fields reinforce both contributions, whereas the two effects partially compensate each other under negative electric fields. Consequently, the magnitude of the critical field reduction depends strongly on the electric field direction.

Another notable feature in Fig. 4a is that the turning point occurs when $E$ exceeds approximately $+0.029$ V/nm, demonstrating suppression of $\mu_{0}H_{\mathrm{C}}$ modulation. Remarkably, complete conversion between the excitonic states $X_{\mathrm{a}}$ and $X_{\mathrm{b}}$ occurs at nearly the same electric field (Figs. 2a-c), where the emission of $X_{\mathrm{b}}$ is quenched by localized states. The near coincidence of these two characteristic electric fields is unlikely to be accidental and suggests that localized states can suppress both the linear ME effect and the electric field tuning of interlayer exchange coupling. Similar behavior is reproduced in Device 2 (Figs. S6 and S12), further supporting the intrinsic nature of this correlation.
  
In summary, we demonstrate deterministic electric field control of excitonic states and magnetic order in bilayer CrSBr. Electric-field-dependent PL measurements reveal both intralayer and interlayer excitons, with the latter exhibiting an unusually small dipole moment of approximately 1 e\AA, indicative of strong interlayer hybridization and a tightly bound character. More importantly, we realize electric-field-driven reversible switching between AFM and FM states and quantitatively attribute the magnetoelectric response to the combined effects of linear magnetoelectric effect and electric-field-modulated interlayer exchange interactions. Our results establish bilayer CrSBr as a platform in which novel excitonic states, interlayer magnetic interactions, and linear magnetoelectric coupling can be simultaneously accessed and controlled by electric fields. These findings advance the understanding of magnetoelectric phenomena in two-dimensional magnetic semiconductors and highlight the potential of CrSBr for electrically programmable spin-optoelectronic and non-volatile memory devices.

\section{Methods}
\subsection{Device fabrication}
Bulk CrSBr crystals were grown by the chemical vapor transport (CVT) method \cite{Qin2026}. Bilayer CrSBr flakes were mechanically exfoliated onto 285 nm $\mathrm{SiO_{2}}$/Si substrates. The layer numbers of CrSBr flakes were determined by optical contrast and confocal Raman spectroscopy (Alpha 300R, WITec). Few-layer graphite and hBN flakes were obtained by mechanical exfoliation. The thicknesses of hBN flakes were determined by an atomic force microscopy (MFP-3D Origin+, Oxford). The dual-gated devices were fabricated using a dry transfer method. A polycarbonate (PC)/polydimethylsiloxane (PDMS) stamp was used to sequentially pick up the graphite, hBN, bilayer CrSBr, hBN, and graphite, after which the whole stack was deposited onto a 285 nm $\mathrm{SiO_{2}}$/Si substrate with pre-patterned electrodes. The entire transfer process was carried out in an argon-filled glovebox. At last, the residual PC on the device surface was dissolved in chloroform.

\subsection{PL spectroscopy}
Magneto-PL and electric-field-dependent PL spectra were measured at 3.7 K in an  closed-cycle cryostat (attoDRY1000) with magnetic fields of up to ±9 T. The magnetic fields were applied along $b$ axis of bilayer CrSBr. The applied electric field was varied by changing the
difference between the top- and bottom-gate voltages. More details on the electric field calculation are provided in Supplementary Note 2. A HeNe laser at 632.8 nm (Thorlabs) was used as the excitation source, with a power of 300 $\mu\mathrm{W}$. The laser polarization was aligned along the $b$ axis of CrSBr to enhance the signal. The laser was focused onto the sample using an objective lens with an NA of 0.68 (Voigt Objective, Attocube), yielding a spot size of approximately 0.6 $\mu\mathrm{m}$. The PL signals were directed into a spectrometer (HRS-300, Princeton Instruments) for dispersion, and analyzed by a charge-coupled device (BLAZE, Princeton Instruments).

\section{Conflict of interest}
The authors declare no conflicts of interest.

\section{Acknowledgement}
The authors acknowledge support by the National Key Research and Development Program of China (2022YFA1602700, 2022YFA1402403, 2021YFA1200800), the National Natural Science Foundation of China (12274155, 52202172, 12574129), the Knowledge Innovation Program of Wuhan-Basic Research (2023010201010040), and the Interdisciplinary program of Wuhan National High Magnetic Field Center (WHMFC2024017), and the Fundamental Research Funds for the Central Universities. Synthesis of boron nitride (K.W. and T.T.) was supported by the Elemental Strategy Initiative conducted by the MEXT, Japan (JPMXP0112101001) and JSPS KAKENHI (JP19H05790, JP20H00354).

\bibliographystyle{apsrev4-2}
\bibliography{main}

\end{document}